\documentclass[aps,preprint,preprintnumbers,nofootinbib,showpacs,superscriptaddress]{revtex4}
\usepackage{graphicx,color,amsmath}
\begin{document}
\title{Possibly New Charmed Baryon States from $\bar B^0\to p\bar p D^{0}$ Decay}
\author{Hai-Yang  Cheng}
\affiliation{Institute of Physics, Academia Sinica, Taipei, Taiwan 115, R.O.C.}
\author{C.Q. Geng}
\affiliation{Department of Physics, National Tsing Hua University, Hsinchu, Taiwan 300, R.O.C.}
\affiliation{Physics Division, National Center for Theoretical Sciences, Hsinchu, Taiwan 300, R.O.C.}
\author{Y.K. Hsiao}
\affiliation{Physics Division, National Center for Theoretical Sciences, Hsinchu, Taiwan 300, R.O.C.}
\date{\today}
\begin{abstract}
We examine the invariant mass spectrum of $D^{0}p$ in $\bar B^0\to p\bar p D^{0}$ decay measured
by BABAR and find that through the 2-step processes of $\bar B^0\to {\bf B_c^+}(\to D^{0} p)\bar p$,
where ${\bf B_c}$ denotes a charmed baryon state, some of the peaks can be identified with
the established $\Sigma_c(2800)^+$, $\Lambda_c(2880)^+$ and $\Lambda_c(2940)^+$.
Moreover, in order to account for the measured spectrum, it is necessary to introduce
a new charmed baryon resonance with $(m,\,\Gamma)=(3212\pm 20,\,167\pm 34)$ MeV.
\end{abstract}

\maketitle
\newpage
{\it Introduction}---
The charmed baryon spectroscopy is expected to be an ideal place
to study the dynamics of the light quarks in the environment of a heavy quark.
Apart from a charm quark, the charmed baryon contains two light quarks as a diquark,
which can be either an anti-triplet or a sextet in the quark model.
Due to various combinations of
individual spins and orbital angular momenta between two light quarks as well as between the
diquark and the charm quark, charmed baryons exhibit a rich spectrum of states~\cite{Migura,Cheng1,Ebert,Chen,CharmB}.
However, so far only a few of charmed baryons have been experimentally identified~\cite{pdg,baryonspectroscopy}.
The $B$ factories as well as the LHCb are able to play an important role in studying charmed baryons.
For example, the charmed baryon states of
$\Lambda_c(2665)^+$, $\Lambda_c(2880)^+$, $\Lambda_c(2940)^+$,
$\Sigma_c(2800)$, $\Xi_c(2980)$, and $\Xi_c(3080)$,
have been observed by
CLEO ~\cite{CLEO1}, BELLE~\cite{BELLE1}, and BABAR~\cite{BABAR1}.
In particular, the decays of $\Lambda_c(2880)^+\to \Sigma_c(2520)^{0(++)}\pi^{+(-)}$ have been
analyzed to assign the quantum number of $\Lambda_c(2880)^+$ to be $J^P=\frac{5}{2}^+$~\cite{BELLE3}.

Recently, the BABAR collaboration  has reported that
both $m_{p\bar p}$ and $m_{D^{0}p}$ invariant mass distributions
in $\bar B^0\to p\bar p D^{0}$
exhibit two sets of peaking data points in the spectra~\cite{BABAR3}.
In the $m_{p\bar p}$ spectrum, the low-mass peaks near the threshold
can be recognized as the threshold effect~\cite{Soni-Hou}.
On the one hand, the exclusion for the range of $m^2_{D^0p}<9$ GeV$^2$
improves the knowledge of the threshold effect, a generic feature
in three-body baryonic $\bar B\to {\bf B\bar B'}M$ decays.
On the other hand, the peaks in the $m_{Dp}$ spectrum
with $m^2_{p\bar p}>5$ GeV$^2$ seem puzzling
as the data points disagree with what expected from the uniform phase-space model.
In this report, we find that, unlike the
peak around the threshold area in the $m_{p\bar p}$ spectrum, those in the $m_{D^{0}p}$ spectrum
{\em cannot} be understood based on the pQCD effect in the creation of the dibaryon~\cite{GengHsiaoCheng,Hsiao}.
We will demonstrate that
the charmed-baryon resonances decaying into $D^{0}p$
fit well to these peaking data points in $\bar B^0\to p\bar p D^{0}$.
It turns out that  one of the resonant charmed baryon states is experimentally not observed yet,
but theoretically studied in the literature~\cite{Migura,Cheng1,Ebert,Chen}.

{\it Formalism}---
To explain the data points in the $m_{D^{0}p}$ invariant mass distribution
in $\bar B^0\to p\bar p D^{0}$, we write down the factorizable
amplitude induced at the tree process $b\to c\bar u d$ \cite{GengHsiaoCheng}
\begin{eqnarray}\label{A-non-r}
{\cal A}(\bar B^0\to p\bar p D^{0})&=&\frac{G_F}{\sqrt 2}V_{cb}V_{ud}^*a_2
\langle D^{0}|(\bar c u)_{V-A}|0\rangle\langle p\bar p|(\bar d b)_{V-A}|\bar B^0\rangle\,,
\end{eqnarray}
where $G_F$ is the Fermi constant,  $V_{cb}$ and $V_{ud}$ represent
the CKM matrix elements for the $b\to c\bar u d$ transition at the
quark level, and $(\bar q_1 q_2)_{V-A}$ stands for $\bar q_1
\gamma_\mu(1-\gamma_5) q_2$. Here, $a_2$ is a parameter which takes
into account the nonfactorizable effects. For the $D$ meson
production, the matrix element in Eq. (\ref{A-non-r}) is given by $\langle D|\bar c\gamma^\mu \gamma_5 u|0\rangle=-if_{D} p^\mu$.
The matrix elements for the $\bar B^0\to p\bar p$ transition are parameterized as \cite{GengHsiaoCheng,GengHsiao3},
\begin{eqnarray}\label{transitionF}
&&\langle p\bar p|\bar d\gamma_\mu b|\bar B^0\rangle=
i\bar u[  g_1\gamma_{\mu}+g_2i\sigma_{\mu\nu}p^\nu +g_3p_{\mu} +g_4 q_\mu +g_5(p_{\bar p}-p_{p})_\mu]\gamma_5v\,,\nonumber\\
&&\langle p\bar p|\bar d\gamma_\mu\gamma_5 b|\bar B^0\rangle=
i\bar u[ f_1\gamma_{\mu}+f_2i\sigma_{\mu\nu}p^\nu +f_3p_{\mu} +f_4 q_\mu +f_5(p_{\bar p}-p_{p})_\mu]        v\,,
\end{eqnarray}
where $p=p_B-p_p-p_{\bar p}$ and $q=p_{p}+p_{\bar p}$ with $p_i\; (i=B,p,\bar{p})$ representing the momenta of the particles,
and the momentum dependences of the form factors are given by
\begin{eqnarray}\label{transitionF1}
f_i=\frac{D_{f_i}}{t^3}\;, \qquad g_i=\frac{D_{g_i}}{t^3}\;,
\end{eqnarray}
with $t\equiv m_{p\bar p}^2$ and $D_{f_i,\,g_i}$ being constants.
Note that Eq. (\ref{transitionF1}) is based on pQCD counting rules \cite{Brodsky1,Hou2}
as three hard gluons are needed to produce ${\bf B\bar B'}$.
Since the $1/t^3$ function peaks at the low mass and flattens out at the large energy region,
the threshold effect in the $m_{p\bar p}$ spectrum~\cite{GengHsiaoCheng} can be easily accommodated.
In the approach of pQCD counting rules, we have explained
the experimental data observed in baryonic $B$ decays, in particular the branching fractions of the following
11 decay modes: $B^-\to p\bar p K^{(*)-}(\pi^-)$,
$\bar B^0\to p\bar p K^{(*)0}$,
$B^-\to \Lambda \bar p \rho^0(\gamma)$, $\bar B^0\to \Lambda \bar p \pi^+$,
$\bar B^0\to n\bar p D^{*+}$, and $\bar B^0\to p\bar p D^{(*)0}$.
Moreover, the predicted ${\cal B}(\bar B\to \Lambda\bar \Lambda \bar K(\pi))$ \cite{LambdaLambdaK},
${\cal B}(\bar B^0\to \Lambda\bar \Lambda D^0)$, and
${\cal B}(B^-\to \Lambda\bar p D^{(*)0})$ \cite{GengHsiaoCheng}
are consistent with the latest measurements \cite{proved}.
These results show that our approach is
a reliable one for tackling the three-body baryonic $B$ decays.
As we shall see, although the amplitude in Eq.~(\ref{A-non-r}) is not suitable to explain the peaking data points in the $m_{D^{0}p}$ spectrum, it
is trustworthy for describing
the non-peaking data points.
Inspired by both experimental~\cite{BABAR1} and theoretical~\cite{molecule} studies
of  $\Lambda_c(2880)^+$ and  $\Lambda_c(2940)^+$ $\to D^0 p$,
we shall examine the $\bar B^0\to p\bar p D^{0}$ decay
via the resonant $\bar B^0\to {\bf B_c}(\to D^{0}p)\bar p $ channels
to reveal
the $m_{D^{0}p}$ invariant mass spectrum.
As in Eq. (\ref{A-non-r}),
via the same effective Hamiltoniam for $b\to c\bar ud$ at the quark level,
the resonant amplitude is derived as
\begin{eqnarray}\label{A-re}
{\cal A_R}(\bar B^0\to ({\bf B_c}\to D^{0}p)\bar  p)
=\frac{G_F}{\sqrt 2}V_{cb}V_{ud}^* a_2
\langle D^{0}p|{\bf B_c}\rangle {\cal R}
\langle {\bf B_c}\bar p| (\bar c u)_{V-A}(\bar d b)_{V-A}|\bar B^0\rangle\;,
\end{eqnarray}
where the factor ${\cal R}$ is the Breit-Wigner factor given by
\begin{eqnarray}
{\cal R}=\frac{i}{q^2-m^2+im\Gamma}\,,
\end{eqnarray}
with $q=p_p+p_{D^{0}}$ and
$m\,(\Gamma)$ the mass (width) of the resonant  $\bf B_c$ state.
It is read
y to evaluate the resonant amplitude ${\cal A_R}(\bar B^0\to ({\bf B_c}\to D^{0}p)\bar  p)$. For example,
if the quantum number of ${\bf B_c}$ is $J^P=1/2^-$,
the combination of the matrix elements for $\bar B^0\to {\bf B_c}\bar p$ and
${\bf B_c}\to D^{0}p$ in Eq. (\ref{A-re}) results in
\begin{eqnarray}
\langle D^{0}p|{\bf B_c}\rangle 
\langle {\bf B_c}\bar p|(\bar c u)_{V-A}(\bar d b)_{V-A}|\bar B^0\rangle=
g\bar u_p \; \Sigma u_{\bf B_c} \bar u_{\bf B_c}(a'+b'\gamma_5)v\;,
\end{eqnarray}
with $\Sigma u_{\bf B_c} \bar u_{\bf B_c}=(\not{\! \!q}+m)$ when the $\bf B_c$ spin is summed over,
where $g$ is the coupling constant for the strong decay ${\bf B_c}\to D^{0}p$.
Without losing generality, this leads to a reduced form of the resonant amplitude
${\cal A_R}(\bar B^0\to ({\bf B_c}\to D^0 p)\bar p)$
 given by
\begin{eqnarray}\label{A-re-2}
{\cal A_R}(\bar B^0\to p\bar p D^{0})=
\frac{G_F}{\sqrt 2}V_{cb}V_{ud}^* \bar u_p
{\cal R}(a+b\gamma_5)v_{\bar p}\;,
\end{eqnarray}
where $(a+b\gamma_5)=a_2 g(\not{\! \!q}+m)(a'+b'\gamma_5)$.
The parameters $a$ and $b$  in the above equation can be treated as constants
for resonances with a narrow width.
Besides, even if $\bf B_c$ carries a higher spin, such as $\Lambda_c(2880)^+$ with spin-5/2,
the spin structure in the propagator can still be summed over and factored into $a$ and $b$.
Since the coupling of ${\bf B_c}\to D^{0}p$ and the parity of ${\bf B_c}$ may not be well determined,
there exist some other possibilities for the amplitude in Eq. (\ref{A-re}),
such as $(a-b\gamma_5)$ and $(b\pm a\gamma_5)$.
By taking $|a|=|b|$, we are able to circumvent this complexity.
Hence, we can obtain the amplitude squared $|\bar A|^2$ with the bar
denoting the summation over  $p$ and $\bar p$ spins.
Using the general partial rate formula for the 3-body decay given by~\cite{pdg}
\begin{eqnarray}\label{int}
d\Gamma=\frac{1}{(2\pi)^3}\frac{|\bar {\cal A}|^2}{32M^3_{\bar B^0}}dm^2_{D^{0}p}\,dm^2_{p\bar p}\,,
\end{eqnarray}
we integrate over $m^2_{p\bar p}$ to obtain
the partial rate as a function of $m_{{D^{0}} p}$ in
$\bar B^0\to p\bar p D^{0}$.

{\it Numerical Analysis}---
We perform a best $\chi^2$ fit to the data.
For the theoretical inputs,
we take the CKM matrix elements
$V_{cb}={A\lambda^2}$ and $V_{ud}=1-\lambda^2/2$  with $A=0.808$ and
$\lambda=0.2253$ from the Particle Data Group~\cite{pdg} along with
the $D$ meson decay constant  $f_D$=0.23 GeV~\cite{pdg,fD(star)}.
For the parameter $a_2$
and the constants $D_{f_i,g_i}$ in Eq. (\ref{transitionF1}),
we adopt their values with errors as in Ref.~\cite{GengHsiaoCheng}, which have been
fitted to explain the total branching fraction of $\bar B^0\to p\bar p D^{0}$.

\begin{figure}[t!]
\centering
\includegraphics[width=2.5in]{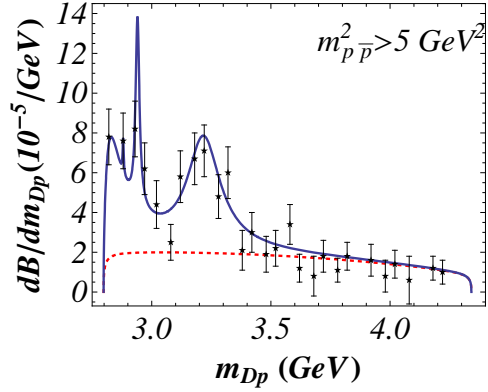}
\caption{Invariant mass spectrum as a function of the invariant mass
$m_{D^0 p}$ in $\bar B^0\to p\bar p D^0$,
where the solid line include both  resonant and non-resonant contributions,
while the dashed line corresponds to the non-resonant contribution only.
}\label{fig-dBdm}
\end{figure}

We now discuss two different scenarios for the data analysis.
In the first scenario, we take the amplitude in Eq. (\ref{A-non-r}) only and
fit the data points in Fig. \ref{fig-dBdm}.
Since the $m_{Dp}$ distribution in Fig. \ref{fig-dBdm}
is shown for $m_{p\bar p}^2>5 \text{GeV}^2$,
we integrate $m_{p\bar p}^2$ in Eq. (\ref{int})
over the range of $5\,\text{GeV}^2< m_{p\bar p}^2 <(m_{p\bar p}^2)_{max}$
with $(m_{p\bar p})_{max}^2$ as a function of $m_{D^{0}p}$ defined in Ref. \cite{pdg}.
The dashed line in Fig.~\ref{fig-dBdm}, drawn to link the non-peaking data points, corresponds to
the amplitude of Eq.~(\ref{A-non-r}) based on pQCD.
%
The value of $\chi^2/d.o.f$ is fitted to be $133/27=4.9$, where $4.9=4.2+0.7$
with the last two numbers coming from the peaking data points of 2.82 - 3.28 GeV 
and non-peaking data points above 3.28 GeV
in the spectrum, respectively.
%
This fitting result strongly suggests the necessity for the presence of resonances.

In the second scenario, we fit the data points in Fig. \ref{fig-dBdm}
by combining the non-resonant amplitude in Eq. ~(\ref{A-non-r})
with the resonant amplitudes for various $\bf B_c$ states
in the form of Eq.~(\ref{A-re}).
It is reasonable to identify  the established $\Sigma_c(2800)^+$, $\Lambda_c(2880)^+$, and $\Lambda_c(2940)^+$,
 with the first three peaks in  the $Dp$ spectrum of $\bar B^0\to p\bar p D^0$.
Particularly, the decays of $\Lambda_c(2880)^+\to D^0 p$ and
$\Lambda_c(2940)^+\to D^0 p$ have been observed~\cite{BABAR1}.
The theoretical inputs of the masses and the decay widths for the established $\bf B_c$ states
are taken from PDG \cite{pdg}, which are also shown  in Table \ref{tab1}.
On the other hand, we need to introduce a new state to accommodate the peak observed around 3.2-3.3 GeV.
The relative phases
among different resonances are assumed to be positive in the fit. Moreover,
the interference between different resonant states is small due to the constrained overlap.
The results are presented in Table \ref{tab1} and
Fig.~\ref{fig-dBdm} where the resonances are accommodated by the solid line.
We note that, by adding the resonant states,
$\chi^2/d.o.f$ is reduced to be 17/21=0.8.
The new resonant state is named as ${\bf B}_c(3212)$ with the fitted value of
$(m,\Gamma)=(3212\pm 20,\,167\pm 34)$ MeV.
In Table \ref{tab1}, as $|a|$ and $|b|$ present the magnitude of a decay,
each central value divided by the error 
indicates the confidence level for a resonant production. 
As a result, the fitting gives the confidence levels of 
5.8$\sigma$, 4.5$\sigma$, 4.6$\sigma$, and 5.0$\sigma$
for the resonant states of $\Sigma_c (2800)^+$, $\Lambda_c (2880)^+$, $\Lambda_c (2940)^+$,
and ${\bf B}_c(3212)^+$,
with the converted  p-values shown in Table \ref{tab1}, respectively.

\begin{table}[h!]
\caption{The resonant charmed baryon (${\bf B_{c}}$)
states in the decay of $\bar B^0\to p\bar p D^0$,
where the values for $|a|$ and $|b|$ are fitting results.
In column 3, input parameters of masses and widths are denoted by the notation $*$.
}\label{tab1}
\begin{tabular}{|c|c|l|c|}
\hline ${\bf B_{c}}$ State &$|a|=|b|$&
$\;\;\;(m,\Gamma)\;$ MeV&p-value\\\hline
$\Sigma_c (2800)^+$         &$4.61\pm 0.79$       &$^*$($2792\pm 14,\;62\pm 60$)          &$3.3\times 10^{-9}$\\
$\Lambda_c (2880)^+$      &$0.09\pm 0.02$        &$^*$($2881.53\pm 0.35,\;5.8\pm 1.1$)&$3.4\times 10^{-6}$\\
$\Lambda_c (2940)^+$      &$0.65\pm 0.14$        &$^*$($2939.3\pm 1.5,\,17\pm 8$)        &$2.1\times 10^{-6}$\\
${\bf B}_c(3212)^+$           &$3.37\pm 0.68$        &\;$(3212\pm 20,\,167\pm 34)$               &$2.9\times 10^{-7}$\\
\hline
\end{tabular}
\end{table}

With the fit mass in the $m_{D^0 p}$ spectrum,
the charmed baryon state ${\bf B_c}(3212)^+$
can be identified with one of the states: $\Sigma_c(3262)$ with $J^P={3/2}^+$ and $\Sigma_c(3268)$ with $J^P={5/2}^+$, or the combination of them. The latter state may arise
from the orbital (2D) excitations of the $c$\,--\,$(ud)$ bound state \cite{Ebert}.
As for the resonant charmed baryon states found in the $\bar B^0\to
p\bar p D^{0}$ decay, one can perform experimental searches.
In analog to the first observation of $\Lambda_c(2940)^+$ in the
process of $e^+ e^-\to \Lambda_c(2940)^+(\to D^{0} p)X$ by
BABAR~\cite{BABAR1}, the ${\bf B}_c$ states can be searched for by scanning the
$D^{0} p$ spectra in ${\bf B_c}\to D^{0}p$ from
the $e^+ e^-$, $p\bar p$, and $pp$ colliders at the $B$ factories,
Tevatron, and LHC, respectively. We also expect that some of the resonant
${\bf B_c}$ states should be observed in the $B$ decays through $\bar
B^0\to {\bf B_c}^+(\to \Sigma_c(2455)^{0,++} \pi^\pm)\bar p$, $\bar
B^0\to {\bf B_c}^+(\to \Sigma_c(2520)^{0,++} \pi^\pm)\bar p$,
 $B^-\to {\bf B_c}^+(\to \Sigma_c(2455)^{0,++} \pi^\pm)\bar
p\pi^-$, and $B^-\to {\bf B_c}^+(\to \Sigma_c(2520)^{0,++} \pi^\pm)\bar
p\pi^-$.
It is important to note that as the resonant $\bar B^0\to p\bar p D^{0}$ decays
imply the existence of the two-body $\bar B^0\to {\bf B_c^+}\bar p$ decays, e.g.
$\bar B^0\to \Lambda_c(2940)^+ \bar p$,
the angular analysis in the $\bar B^0\to {\bf B_c}\bar p$ decays
will be very useful to extract the right quantum numbers for ${\bf B}_c$ states.
Note that the spin of $\Lambda_c(2940)$ is still unknown.
Finally,  we remark that the charmed baryons $\Sigma_c$ and $\Lambda_c$
in our study would have the counterparts with the well measured
charm-strange baryons $\Xi_c$~\cite{CharmB,Liu} based on the $SU(3)$
flavor symmetry.

{\it Conclusions}---
We have analyzed the
peaks in the $m_{D^{0}p}$ spectrum
observed  in the $\bar B^0\to p\bar p D^{0}$ decay with  $m_{p\bar p}^2>5\,\text{GeV}^2$
and found that some of the peaking points can be recognized as the charmed baryon resonances
in $\bar B^0\to {\bf B_c}(\to D^{0} p)\bar p$ channels.
In addition to the established $\Sigma_c(2800)^+$, $\Lambda_c(2880)^+$, and $\Lambda_c(2940)^+$,
a new resonance named ${\bf B}_c(3212)^+$ has been introduced, which is
fitted to have $(m,\,\Gamma)=(3212\pm 20,\,167\pm 34)$ MeV
for $\bar B^0\to p\bar p D^0$.
Moreover, since the resonant $\bar B^0\to {\bf B_c}(\to D^{0} p)\bar p$ decays are
the consequences of the two-body $\bar B^0\to {\bf B_c} \bar p$ decays,
the studies of the angular distributions in the two-body modes
will help determine the spins for the new charmed baryon states,
such as, $\Lambda_c(2940)$ and ${\bf B_c}(3212)$.

{\it Note added}---
After we posted the paper to the e-print archive (arXiv:1205.0117 [hep-ph]), the BABAR collaboration has presented
the study of $B^-\to \Sigma_c^{++}\bar p\pi^-\pi^-$ \cite{BABAR4}, where
they have observed a new resonance ${\bf B_c}(3245)^0$ 
with $(m,\Gamma)=(3245\pm 20,108\pm 6)$ MeV in the spectrum of $\Sigma_c^{++}\pi^-\pi^-$.
With similar masses and decay widths,
${\bf B}_c(3245)^0$ and ${\bf B}_c(3212)^+$ can be regarded as
the neutral and charged-one components of the excited baryon states of
$(\Sigma_c^{++},\Sigma_c^+,\Sigma_c^0)$ predicted in Ref. \cite{Ebert}.

{\it Acknowledgments}---
We would like to thank Dr. Yubing Dong for useful discussions.
The work was supported in part by National Center of Theoretical Science,
and  National Science Council of R.O.C. under
Grants Nos. NSC-100-2112-M-001-009-MY3 and NSC-98-2112-M-007-008-MY3.

\end{document}